\documentclass[prb,twocolumn,showpacs,preprintnumbers,amsmath,aps]{revtex4}
\usepackage{graphicx,hyperref}
\usepackage{amssymb}
\usepackage{bm}
\usepackage{subfigure}

\begin{document}

\title{Random-field-like criticality in glass-forming liquids}

\author{Giulio Biroli} \email{giulio.biroli@cea.fr}
\affiliation{IPhT, CEA/DSM-CNRS/URA 2306, CEA Saclay, F-91191 Gif-sur-Yvette Cedex, France}

\author{Chiara Cammarota} \email{chiara.cammarota@roma1.infn.it}
\affiliation{IPhT, CEA/DSM-CNRS/URA 2306, CEA Saclay, F-91191 Gif-sur-Yvette Cedex, France}
\affiliation{Dipartimento di Fisica, Ed. Marconi, "Sapienza" Universit\`a di Roma, P.le A. Moro 5, 00185 Roma Italy}

\author{Gilles Tarjus} \email{tarjus@lptmc.jussieu.fr}
\affiliation{LPTMC, CNRS-UMR 7600, Universit\'e Pierre et Marie Curie,
bo\^ite 121, 4 Pl. Jussieu, 75252 Paris c\'edex 05, France}

\author{Marco Tarzia} \email{tarzia@lptmc.jussieu.fr}
\affiliation{LPTMC, CNRS-UMR 7600, Universit\'e Pierre et Marie Curie,
bo\^ite 121, 4 Pl. Jussieu, 75252 Paris c\'edex 05, France}

\date{\today}

\begin{abstract}
We introduce an approach to derive an effective scalar field theory for the glass transition; the fluctuating field 
is the overlap between equilibrium configurations. We apply it to the case of constrained liquids for which the introduction 
of a conjugate source to the overlap field was predicted to lead to an equilibrium critical point. We show that the long-distance 
physics in the vicinity of this critical point is in the same universality class as that of a paradigmatic disordered model: the random-field 
Ising model. The quenched disorder is provided here by a reference equilibrium liquid configuration. We discuss to what extent this 
field-theoretical description and the mapping to the random field Ising model hold in the whole supercooled liquid regime, in particular 
near the glass transition.  
\end{abstract}

\maketitle

\label{introduction}
One of the first steps in the analysis of a standard phase transition consists in identifying the correct order parameter. Once this crucial step is done, 
one can first construct a Landau functional and analyze it to obtain a mean-field description;  then, eventually, one 
can promote the order parameter to a truly fluctuating field and study the associated field theory in order to get a full-fledged theoretical description. In 
the case of the glass transition establishing what is the correct order parameter is by no means an easy task.
The Random First-Order Transition (RFOT) theory\cite{KTW,Wolynesbook} identifies as order parameter the similarity, also called overlap, 
between equilibrium configurations. More specifically, take an equilibrium (reference) configuration and then restrict the thermodynamic sampling 
only to equilibrium configurations constrained to have a high overlap with the reference one on the boundary of the sample. By definition, a RFOT 
takes place when this boundary condition forces the entire system to have a large value of the overlap instead of the zero (or very low) one 
characteristic of the liquid phase. This is analogous to forcing a positive boundary magnetization in the case of a ferromagnetic phase transition. The 
identification of this order parameter was first made in the context of disordered mean-field systems \cite{franzparisi-potential} but it was soon realized 
that the overlap and its fluctuations provide an interesting tool in the study of glassy systems, in particular supercooled liquids, irrespective 
of the presence or not of quenched disorder\cite{franzparisi-potential2}.
Studying the response to perturbations directly acting on the order parameter, as done in usual phase transitions, has recently allowed an access to the 
growing static length accompanying the slowing down of the dynamics. 
In the RFOT context the perturbation is a pinning field that forces configurations to have a high overlap locally; the corresponding correlation 
length is called ``point-to-set".\cite{BB} As discussed before, promoting the order parameter to a fully fluctuating field is a way to study 
fluctuations and correlations beyond mean-field theory: this was recently done to analyze the slow but intermediate $\beta$ relaxation 
in the vicinity of the ``ideal'' dynamical transition that is found both in the mode-coupling approach and in the RFOT theory of liquids.\cite{parisi,franzjacquin} 
In particular, it has been shown that the dynamical transition is in the universality class of the spinodal 
of the random-field Ising model (RFIM). Both singularities of course can only be present when activated events such as nucleation are not taken 
into account.
Finally, the overlap has also been the focus of an intense numerical research 
in model supercooled liquids: the distribution of the fluctuations of the uniform overlap between equilibrium configurations has been computed and 
found to develop a nontrivial, non-Gaussian, shape as one cools the liquid.\cite{FP-V,cammarotaetal-V,berthier-V}\\
The aim of our work is to develop an effective field theory of glass-forming systems directly formulated in terms of an overlap field. This is 
highly desirable for several reasons: first, it allows one to focus directly on what is thought to be the physically relevant field; second, it leads to 
a scalar field theory in the presence of quenched disorder and should therefore settle the recurrent debate about whether the 
glass transition is related to random-field, random-bond or spin-glass physics.\cite{KTW,WolynesWal,moore}
Our approach is able to capture nonperturbative effects conjectured to be crucial to describe the glass transition: by perturbatively integrating 
out irrelevant degrees of freedom only, we derive an effective theory for the relevant field, identified as the the overlap with a 
reference equilibrium configuration. The model (or field theory) 
obtained by this procedure can then be nonperturbatively studied either by renormalization-group analysis or by computer simulations. \\
In the following we first introduce our method in a general setting for glass-forming liquids. Then, we apply it to the \textit{critical 
point} that terminates the transition line in an extended phase diagram where one introduces a coupling between liquid configurations  and 
we show that the critical behavior is the same as in the equilibrium RFIM. The motivation for studying this specific region of parameters,
besides providing a first progress towards a comprehensive field theory of the glass transition, stems from recent numerical 
works \cite{FP-V,cammarotaetal-V,berthier-V} that have directly focused on the behavior of supercooled liquids in the presence of such an attractive coupling 
and have provided evidence for a first-order transition line and a terminal critical point. In consequence, our predictions 
are prone to direct tests in the future. 
(Note that the analysis of Ref. [\onlinecite{FP-CP1}] does not apply here since, in the presence of a nonzero coupling, the first-order transition 
line is \textit{not} a RFOT and the spinodal is \textit{not} equivalent to a dynamical mode-coupling-like transition.\cite{CB-JCP})

Consider a glass-forming liquid formed by $N$ particles and described by a Hamiltonian $H[\bf{r}^N]$ where $\bf{r}^N$ denotes a configuration 
of the $N$ particles. 
We consider a reference equilibrium configuration $\bf{r}_0^N$ and define the overlap at 
point $x$ between the latter and another configuration $\bf{r}^N$ as $\hat q_x[\bf{r}^N,\bf{r}_0^N]:=\int_y f(y) 
[\hat \rho(x+\frac{y}{2} \vert\bf{r}^N )\hat \rho(x-\frac{y}{2} \vert\bf{r}_0^N )-\rho^2]$, where $\int_y\equiv \int d^D y$ and $f(y)$ is a 
smoothing function of short range (typically the cage size); 
$\rho(x \vert\bf{r}^N )=\sum_{i=1}^N\delta^{(D)}(x-r_i)$ is the microscopic density at point $x$ and $\rho$ is the mean liquid density. We can 
now define an overlap field $p(x)$ and introduce an effective Hamiltonian or action for this field,
\begin{equation}
 \label{eq_disordered_action}
 S[p\vert {\bf{r}_0^N}]=-\log \int \frac{d{\bf{r}^N}}{N!} \delta[p-\hat q[{\bf{r}^N},{\bf{r}_0^N}]] e^{-\beta H[\bf{r}^N]}
\end{equation}
where $\delta[\;]$ is a functional that enforces a delta function at each point $x$ and $\beta=1/(k_B T)$. 
The probability to observe a certain profile of the density field is given by 
$\exp(- S[p\vert \bf{r}_0^N])$. Thermodynamic quantities and 
correlation functions of the overlap field, e.g. point-to-set ones, are obtained as usual from a ``partition function'' and the associated functional $W$, 
\begin{equation}
 \label{eq_disordered_generating_functional}
 e^{W[\epsilon \vert {\bf{r}_0^N]}}= \int \mathcal{D} p\, e^{-S[p\vert {\bf{r}_0^N}]+\int d^Dx  \epsilon(x) p(x)} \, ,
\end{equation}
where we have introduced an auxiliary coupling $\epsilon(x)$ that, in field-theoretical language, plays the role of a ``source'' for generating 
the connected correlation functions of the overlap field. A RFOT corresponds to the appearance for $\epsilon=0$ of long-range order in the 
overlap field, which acquires a large value in the entire sample.
Because of the reference configuration, $\bf{r}_0^N$, the action $S$ describes a scalar field
theory in the presence of quenched disorder. In order to analyze it and understand in more detail what kind of 
disorder is generated by $\bf{r}_0^N$ one can study the cumulants of $S$ by considering replicas of the original system. As known 
in the context of the critical behavior of the RFIM,\cite{tarjus-tissier,tissier-tarjus} (see also Refs. [\onlinecite{Dzero-Schmalian-Wolynes,parisi}])
$\exp(-S_{rep}[\{p_a\}])=\overline{\exp(-\sum_{a=1}^n  S[p_a\vert \bf{r}_0^N])}$ generates the cumulants of the action 
$S[p\vert \bf{r}_0^N]$ through an expansion in increasing number of free 
replica sums: \cite{tarjus-tissier,tissier-tarjus}
\begin{equation}
\begin{aligned} 
\label{eq_freereplicasums_S}
S_{rep}[\{p_a\}]=&\sum_{a=1}^n S_1[p_a] - \frac{1}{2} \sum_{a,b=1}^nS_2[p_a,p_b] \\& +
 \frac{1}{3!} \sum_{abc=1}^nS_3[p_a,p_b,p_c] + \cdots
\end{aligned}
\end{equation}
where $S_l[p_1,\cdots,p_l]$ is the $l$th cumulant: {\it e.g.}, $\overline{S[p\vert {\bf{r}_0^N]}}=S_1[p]$ and 
$\overline{S[p_1\vert {\bf{r}_0^N}]S[p_2\vert {\bf{r}_0^N}]}-\overline{S[p_1\vert \bf{r}_0^N]}\,\overline{S[p_2\vert \bf{r}_0^N]}=S_2[p_1,p_2]$. 
The Franz-Parisi potential,\cite{franzparisi-potential} which is the average free-energy cost to keep two configurations at a fixed global overlap,  is the 
Legendre transform of the first cumulant of $W[\epsilon \vert \bf{r}_0^N]$ (and is not equal to $S_1$ except in the mean-field limit).\\
Our goal is to derive the action for the overlap field and its cumulants in glass-forming systems. 
To proceed, one can formally rewrite
\begin{equation}
\begin{aligned}
 \label{eq_replica_action}
e^{-S_{rep}[\{p_a\}]}\propto & \int d{\bf{r}_0^N}  \prod_{a=1}^n d{\bf{r}_a^N} 
\delta[p_a-\hat q[{\bf{r}_a^N,\bf{r}_0^N}]] \\& \times e^{-\beta ( H[{\bf{r}_0^N}]+\sum_{a=1}^n H[{\bf{r}_a^N}])} \, .
\end{aligned}
\end{equation} 
The $n$ replicas plus the reference configuration can now be described by Greek letters $\alpha=0,1,\cdots,n$ whereas  Roman 
letters are still used for replicas from $1$ to $n$ only. We also introduce additional collective fields $q_{\alpha\beta}(x)$ that describe the 
overlap between two different replicas $\alpha$ and $\beta$. Eq.~(\ref{eq_replica_action})  then becomes
\begin{equation}
\begin{aligned}
 \label{eq_replica_action2}
e^{-S_{rep}[\{p_a\}]} &\propto \int  \prod_{ab \neq} \mathcal D q_{ab}\bigg [\int \prod_{\alpha}  d {\bf{r}_{\alpha}^N}
\prod_{\alpha\beta \neq} \delta[q_{\alpha\beta}-\hat q[{\bf{r}_{\alpha}^N,\bf{r}_{\beta}^N}]]
\\& \times e^{-\beta\sum_{\alpha} H[{\bf{r}_{\alpha}^N}]}\bigg ] \propto \int  \prod_{ab \neq} \mathcal D q_{ab} \, e^{- \mathcal S[\{p_a,q_{ab}\}]}
\end{aligned}
\end{equation}
where we have used the notation $q_{0a}=q_{a0}\equiv p_a$ and $S[\{p_a,q_{ab}\}]$
is defined as minus the logarithm of the expression between parethensis.
Our approach differs from the usual replica one in that it treats the fields $p_a=q_{0a}$ and $q_{ab}$ differently. The underlying working 
hypothesis, which can at least be checked in the vicinity of the terminal critical point in the presence of a nonzero coupling $\epsilon$, is 
that the $p_a$'s may develop long-range fluctuations while the $q_{ab}$'s are harmless for the long-distance physics and can be approximately 
integrated out.  
In the following we show how this procedure can be carried out near the terminal critical point.\\
One first needs a tractable expression for the action $\mathcal S[\{p_a,q_{ab}\}]$ in terms of the overlap fields [defined via 
Eq.~(\ref{eq_replica_action2})]. This action of course depends on the microscopic details of the glass-forming system under study 
and deriving its expression can be a rather formidable task. 
One can formally derive $\mathcal S[\{q_{\alpha\beta}\}]$ from the Morita-Hiroike functional of the 
1- and 2-particle densities\cite{morita-hiroike} of the replicated $(n+1)$-component liquid mixture along lines similar to those 
followed by Ref.~[\onlinecite{franzjacquin}]. A short-cut is however provided by the coarse-grained  effective Landau-like functional considered in 
Ref.~[\onlinecite{Dzero-Schmalian-Wolynes}]:
\begin{equation}
\begin{aligned}
 \label{eq_replica_action_wolynes}
\mathcal S[\{q_{\alpha\beta}\}]=&\frac{E_0}{k_BT} \int_x \bigg \{\frac{c}{2} \sum_{\alpha\beta \neq} (\partial q_{\alpha\beta}(x))^2
 + \sum_{\alpha\beta \neq} V(q_{\alpha\beta}(x)) \\& -\frac{u}{3} \sum_{\alpha\beta\gamma \neq}
q_{\alpha\beta}(x)q_{\beta\gamma}(x)q_{\gamma\alpha}(x)\bigg \},
\end{aligned}
\end{equation}
where $V(q)=(t/2)q^2-[(u+w)/3]q^3+(y/4)q^4$ and the primary dependence on temperature is given by $t\approx k_B(T-T_0)/E_0$ with $E_0$ 
the typical energy scale of the liquid and $T_0$ a constant with dimension of temperature. 
In numerical applications, we focus on the parameter values found to roughly reproduce properties of glass-forming liquid
ortho-terphenyl: $u=0.385, w=2.73, y=1.82$, and $c=1$ (in appropriate length units).\cite{Dzero-Schmalian-Wolynes}\\
To derive an effective field theory for the overlaps $p_a(x)$ with a reference configuration, one needs to perform the functional integration 
over the $q_{ab}(x)$'s while keeping the fields $p_a(x)$ fixed [see Eq.~(\ref{eq_replica_action2})]. It is clear from Eq. (\ref{eq_replica_action_wolynes}) 
that, as discussed above, nonzero $p_a$'s exert an external source or field on the $q_{ab}$'s. In Eq.~(\ref{eq_replica_action_wolynes}) 
the cubic term generates a contribution $-u\sum_{ab \neq}p_a(x)p_b(x)q_{ab}(x)$. In consequence, the $q_{ab}$'s do not develop fluctuations 
on all scales and stay ``massive''  near the terminal critical point.\cite{footnote}  
It is then sufficient to perform the functional integration on these fields through a perturbative treatment. Perturbation 
is carried out with a saddle-point approximation as zeroth order. 
The saddle-point equations for the $q_{ab}$'s with $a\neq b$ read
\begin{equation}
\begin{aligned}
\label{eq_saddle-point}
c \,\partial^2q_{ab*}(x) &+ V'(q_{ab*}(x))=u p_a(x) p_b(x) \\&+ 
u \sum_{c \neq a ,b} q_{ac*}(x)q_{cb^*}(x)\, .
\end{aligned}
\end{equation}
The solution of this equation has to be inserted back into the action in order to obtain the final result. Since we are interested in the long 
wave-length fluctuations of the $p_a$'s it is sufficient to solve the saddle-point equations in an expansion in the gradient term, the zeroth order then corresponding 
to simply neglecting the gradient.\\
All quantities can be expanded in increasing number of free replica sums as in Eq.~(\ref{eq_freereplicasums_S}), \textit{e.g.},
\begin{equation}
 \label{eq_saddle-point_qab}
\begin{aligned} 
q_{ab*}(x)=q^{[0]}_x[p_a,p_b]+\sum_c q^{[1]}_x[p_a,p_b\vert p_c]+\mathcal O(\sum_{cd}) \,.
\end{aligned}
\end{equation}
Such expansions allow algebraic manipulations that lead to well defined and unique expressions of 
the various orders \cite{tarjus-tissier,tissier-tarjus, ledoussal-wiese}. In the lowest order in the gradient amplitude parameter $c$, 
Eqs.~(\ref{eq_saddle-point}) and (\ref{eq_saddle-point_qab}) lead for instance to $q^{[0]}_x[p_a,p_b]\equiv q^{[0]}(p_a(x),p_b(x))$ and
\begin{eqnarray}
 \label{eq_saddle-point_qzero}
&V'(q^{[0]}(p_a,p_b))&=u p_a p_b \\
&&-u q^{[0]}(p_a,p_b)[q^{[0]}(p_a,p_a)+q^{[0]}(p_b,p_b)]\, .\nonumber
\end{eqnarray}
With the above results, one immediately derives from Eqs.~(\ref{eq_freereplicasums_S}) and (\ref{eq_replica_action2}) the expressions 
of the first two cumulants of the action for the $p_a$'s at the level of the saddle-point approximation and including 
$\mathcal O(\partial^2)$ terms only:
\begin{equation}
 \label{eq_saddle-point_S1zero}
\begin{aligned} 
& S_1[p_a]=\frac{E_0}{k_BT}\int_x \bigg\{c (\partial p_a)^2-\frac 1 2 c (\partial q^{[0]}(p_a,p_a))^2
+2V(p_a)\\&-V(q^{[0]}(p_a,p_a))-\frac{2u}{3}q^{[0]}(p_a,p_a)^3+u p_a^2q^{[0]}(p_a,p_a)
\bigg\}
\end{aligned}
\end{equation}
and
\begin{equation}
 \label{eq_saddle-point_S2zero}
\begin{aligned} 
&S_2[p_a,p_b]=-\frac{E_0}{k_BT}\int_x \bigg\{\frac c 2 (\partial q^{[0]}(p_a,p_b))^2-2 u p_a p_b q^{[0]}(p_a,p_b)
 \\& + u \left[q^{[0]}(p_a,p_a)+q^{[0]}(p_b,p_b)\right]q^{[0]}(p_a,p_b)^2+ 2 V(q^{[0]}(p_a,p_b))\bigg\}
\end{aligned}
\end{equation}
where the explicit $x$-dependence has been omitted. This derivation is easily extended to the higher 
orders but the algebra rapidly becomes tedious and the results are given in the supplementary information (SI).\\
The above cumulants describe a scalar field theory for a disordered system.\cite{tarjus-tissier,tissier-tarjus,ledoussal-wiese,parisi} 
We now more specifically consider the vicinity of the terminal critical point in the $(T,\epsilon)$ plane. At the saddle-point level, the 
critical point is defined by 
the following conditions:
\begin{equation}
 \label{eq_critical_point}
\begin{aligned} 
\frac{\partial S_1(p_1)}{\partial p_1}\bigg \vert_c=\epsilon_c\, , \frac{\partial^2 S_1(p_1)}{\partial p_1^2}\bigg \vert_c=0\, ,
 \frac{\partial^3 S_1(p_1)}{\partial p_1^3}\bigg \vert_c=0 \, .
\end{aligned}
\end{equation}
When inserted in Eq.~(\ref{eq_saddle-point_S1zero}), this gives $\epsilon_c=0.602 \,E_0/(k_B T_c)$, $t_c=1.680$, $p_c=0.534$ and 
$q^{[0]}_c=0.072$ (using the relation\cite{Dzero-Schmalian-Wolynes} between $t$ and $T$ one finds $E_0/(k_B T_c)=0.952$).  
One can then expand the cumulants $S_l[p_1,\cdots,p_l]$ around the critical value after defining 
$\phi_a(x)=p_a(x)-p_c$.  Keeping only terms to order $\phi^4$ (higher-order terms are expected to be irrelevant at criticality) leads to a 
Wilson-Ginzburg-Landau action for the replica scalar fields $\phi_a(x)$: 
\begin{equation}
 \label{eq_WGLreplica_action}
\begin{aligned} 
&S_{rep}[\{\phi_a\}]-\epsilon_c \int_x  \phi_a(x)=\int_x \bigg\{\sum_a \bigg[c' (\partial \phi_a)^2 + \frac{r_2}{2} \phi_a^2 
\\&+ \frac{r_3}{3!} \phi_a^3+ \frac{r_4}{4!} \phi_a^4\bigg ]-
\frac 12 \sum_{ab}\phi_a\phi_b
 \bigg [\Delta_{20}+\frac{\Delta_{21}}{2}(\phi_a+\phi_b) +
 \\&\frac{\Delta_{22}}{4}\phi_a\phi_b+
\frac{\Delta_{23}}{2}(\phi_a^2+\phi_b^2)\bigg ]
+\frac{1}{3!} \sum_{abc}\phi_a\phi_b\phi_c\bigg[\Delta_{30}+
\frac{\Delta_{31}}{3}\\&
\times(\phi_a+\phi_b+\phi_c)\bigg ]
-\frac{\Delta_{40}}{4!} \sum_{abcd}\phi_a\phi_b\phi_c\phi_d +\mathcal O(\sum_{abcde}) \bigg\}
\end{aligned}
\end{equation}
where  $r_2$ and $r_3$ are zero at the saddle-point (mean-field) critical point and all other coefficients are evaluated at the latter point: $c'=0.91$,
$r_4=23.33,\Delta_{20}=0.22,\Delta_{21}=0.47,\Delta_{22}=1.23,\Delta_{23}=0.04,\Delta_{30}=-0.12,\Delta_{31}=-0.51,
\Delta_{40}=0.33$. We have neglected some square gradient terms of higher orders in the fields because they are either less 
relevant in the RG sense or are anyhow generated along the RG flow. Beyond the precise values, the important fact is that $r_4$, 
$\Delta_{20}$, $\Delta_{22}$, and $\Delta_{40}$ are greater than zero. It is then easily inferred that the above replicated action is obtained 
from a random one, $S[\phi\vert h,\delta r_2,\delta r_3]$, which describes the large-scale fluctuations of the overlap field $p(x)$ with 
a reference equilibrium configuration:
\begin{equation}
 \label{eq_WGLdisorder_action}
\begin{aligned} 
&S[\phi\vert h,\delta r_2,\delta r_3]=\int_x \bigg\{c' [\partial \phi(x)]^2 + \frac{r_2}{2} \phi(x)^2+ \frac{r_3}{3!} \phi(x)^3+\\
&\frac{r_4}{4!} \phi(x)^4\bigg\} +\int_x \bigg\{\frac{\delta r_2(x)}{2} \phi(x)^2 + \frac{\delta r_3(x)}{3!} \phi(x)^3- h(x)\phi(x)\bigg\}
\nonumber
\end{aligned}
\end{equation}
where the random field $h(x)$, the random mass $\delta r_2(x)$, the random cubic coupling $\delta r_3(x)$ are delta-correlated processes 
with zero mean and higher cumulants given by
\begin{equation}
 \label{eq_disorder_cumulants}
\begin{aligned} 
&\overline{h(x)h(x')}=\delta_{xx'}\Delta_{20}\, , \overline{h(x)h(x')h(x'')}=\delta_{xx'x''}\Delta_{30}\, , \\&
\overline{h(x)h(x')h(x'')h(x''')}\vert_{cum}=\delta_{xx'x''x'''}\Delta_{40}\, ,\\&
\overline{\delta r_2(x)\delta r_2(x')}=\delta_{xx'}\Delta_{22}\, , \overline{h(x)\delta r_2(x')}=-\delta_{xx'}\Delta_{21}\, ,\\&
\overline{h(x)h(x')\delta r_2(x'')}=\delta_{xx'x''}\Delta_{31}\, , \overline{h(x)\delta r_3(x')}=-\delta_{xx'}\Delta_{23}\, ,\nonumber
\end{aligned}
\end{equation}
etc, with $\delta_{xx'}, \delta_{xx'x''}, \cdots$  short-hand notations for delta functions and products of delta functions. 
The resulting theory is thus a $\phi^4$ one in the presence of quenched disorder without statistical inversion ($Z_2$) symmetry. As is known 
from the theory of disordered systems,\cite{nattermann} the random field is the most relevant of the above terms at criticality and it leads to 
a universality class controlled by a nontrivial zero-temperature fixed point. This is so even in the absence of $Z_2$ symmetry, as can be found 
from a simple Harris-like criterion and has recently been confirmed by a full nonperturbative RG analysis.\cite{Balog-Tarjus-Tissier}\\
The conclusion of our analysis is that the terminal critical point found in the presence of a conjugate source within the mean-field theory belongs 
in finite dimension to the universality class of the RFIM. Of course this is valid {\it if the transition is not destroyed by the disorder}. One therefore 
needs to compare the ``bare" strength of the random field, $\sqrt \Delta$, obtained from $S_2$ to that of the surface tension $Y$ obtained from $S_1$. 
We have computed the latter as the free energy cost per unit surface between two regions with high and low overlap\cite{langer,franz} at coexistence 
far below the (mean-field) critical point and evaluated $\Delta$ at the same temperature, see SI for more details. 
The output is $\sqrt{\Delta}/Y\simeq 0.47$, which from known numerical results on 
the RFIM in $d=3$ (see also SI) is compatible with the existence of a transition.\cite{nattermann} 
We have repeated the whole analysis for the finite dimensional $3$-spin model with weak long-range interactions (see SI) and 
obtained that the mapping to the RFIM also holds. However, in this case, we have found $\sqrt{\Delta}/Y\simeq 2.24$, a value likely too large 
for a transition to survive in $d=3$, in agreement with results and arguments presented in Refs. [\onlinecite{CBTT,moore2}]. \\
The Landau-like functional for glass-forming liquids\cite{Dzero-Schmalian-Wolynes,WolynesWal} used as initial input in the present 
field-theoretical approach is just a crude approximation. Deriving from first principles a proper starting point with effective parameters 
that incorporate the microscopic information about glass-formers is then a crucial task. As illustrated by recent numerical 
work,\cite{berthier-V,cammarotaetal-V} this is now within reach.  Indeed, by constraining the overlap to a fixed value in 
a \textit{small-size system} (less than the point-to-set length), it should be possible to measure 
the local part of the first cumulants $S_1$ and $S_2$ (see above), thus allowing a direct evaluation of the \textit{bare magnitude} of the different sources 
of quenched disorder in model supercooled liquids.\\
An important question for the theory of glass-forming systems is to what extent the mapping to the RFIM found near the terminal critical point 
in the presence of a nonzero coupling $\epsilon$ is general? We have verified that it also holds 
for the transition line in the temperature-$\epsilon>0$ plane, which is therefore a first-order transition in the presence of a random field. 
Results of Refs. [\onlinecite{FP-CP1,CB-JCP}] suggest that it also applies for the continuous glass transition taking place 
at the terminal point in the phase diagram of pinned systems \cite{CB-PNAS}.
The more interesting and physical case corresponds to the situation in the absence of coupling $\epsilon$, 
where an ideal glass transition of RFOT type is predicted at the mean-field level. Although the mapping 
then holds for completely connected models (one can \textit{e.g.} show that the Random Energy Model\cite{REM} 
maps exactly to a zero-dimensional RFIM), its extension to finite-dimensional 
systems is not straightforward and it will be the focus of a future publication. Whether the physics of the glass transition is related to the 
RFIM \cite{KTW,WolynesWal}, to a spin glass in a field \cite{moore} or even to a different kind of universality class still remains 
an open question, but its answer seems now within reach. 

\begin{acknowledgments}
After completion of our work we came to know that S. Franz and G. Parisi have also addressed the problem of the critical point of 
constrained glassy systems (arXiv:1307.4955). Their approach is different but leads to the same conclusion regarding the universality 
class of the terminal point in the T-$\epsilon$ plane.We acknowledge support from the ERC grants NPRGGLASS and (CC) CRIPHERASY (no. 247328).  
\end{acknowledgments}

\clearpage
\begin{center}
\bf{\Large Supplementary information}
\end{center}
\section{Calculation of the cumulants of the effective action for the overlap field}

As explained in the main text, the central quantity is the effective Hamiltonian or action $S[p\vert \bf{r}_0^N]$ describing the configurations of a 
glass-forming system with an overlap $p(x)$ with a reference equilibrium configuration denoted $\bf{r}_0^N$. Our aim is to compute the cumulants 
of this action, which are generated according to
\begin{equation}
\label{eq_cumulants} 
\exp(-S_{rep}[\{p_a\}])=\overline{\exp(-\sum_{a=1}^n  S[p_a\vert \bf{r}_0^N])}
\end{equation}
through an expansion in increasing number of free replica sums of the ``replicated action'' $S_{rep}[\{p_a\}]$:
\begin{equation}
\label{eq_freereplicasums_S_app}
S_{rep}[\{p_a\}]=\sum_{l=1}^{\infty} \frac{(-1)^{l-1}}{l!}\sum_{a_1\cdots a_l=1}^n S_l[p_{a_1},\cdots,p_{a_l}] \,,
\end{equation}
where $S_l[p_{1},\cdots,p_{l}]=\overline{S[p_1\vert \mathbf{r}_0^N] \cdots S[p_l\vert \mathbf{r}_0^N]}\big \vert_{cum}$.

For the sake of concreteness, we show the details of the calculation for the $p$-spin models in the Kac limit\cite{franz,franz05} with $p\geq 3$ [$p$ 
should not be confused with the overlap field $p(x)$]. All the steps of the derivation can be similarly repeated for the Landau-like functional  
given in Eq.~(6) of the main text. The action $\mathcal S[\{p_a,q_{ab}\}]$ that appears in Eq.~(5) of the main text and 
is the starting point of the computations can then be obtained by standard methods and reads in the case of the $p$-spin Kac model\cite{franz,franz05}
\begin{equation}
\begin{aligned}
 \label{eq_replica_action_pspin}
\mathcal S[\{q_{\alpha\beta}\}]=& \int_x \bigg \{\frac{c}{2} \sum_{\alpha\beta \neq} (\partial q_{\alpha\beta}(x))^2
-\frac{\beta^2}{4} \sum_{\alpha\beta \neq} q_{\alpha\beta}(x)^p\\&  - \frac 12 \mathrm{Tr} \log[\boldmath{I} 
+\boldmath{U}(\{q_{\alpha\beta}(x)\})]
\bigg \},
\end{aligned}
\end{equation}
where $\boldmath{I}$ is the identity and $\boldmath{U}$ an $(n+1)\times (n+1)$ matrix with all diagonal elements equal to $0$, 
$U_{0a}=q_{0a}=p_a$, $U_{a0}=q_{a0}=p_a$, and $U_{ab}=q_{ab}$ for $a \neq b$. (we recall that Greek letters are use for the 
$n+1$ copies of the original system, including the reference configuration $\alpha=0$, whereas Latin ones are reserved for the $n$ 
replicas other than the reference one.).\cite{footnote0}
The action for the overlaps $p_a$ with the reference configuration is then obtained by integrating 
out the overlaps $q_{ab}$:
\begin{equation}
\begin{aligned}
 \label{eq_replica_action2}
e^{-S_{rep}[\{p_a\}]} \propto \int  \prod_{ab \neq} \mathcal D q_{ab} \, e^{- \mathcal S[\{p_a,q_{ab}\}]} \,.
\end{aligned}
\end{equation}
In the Kac limit, the gradient term in Eq. (\ref{eq_replica_action_pspin}) can be considered as a perturbation and the integral over the 
$q_{ab}$'s can be computed through a saddle-point approximation. In the following we therefore only detail the computation of the 
local part of $S_{rep}[\{p_a\}]$ that is obtained via the saddle-point. The gradient terms can then be trivially added.

When dropping the gradient term and considering spatially uniform overlaps $p_a$'s, the saddle-point equations for the $q_{ab}$'s 
with $a\neq b$ read
\begin{equation}
\label{eq_saddle-point}
p\frac{\beta^2}{4} q_{ab*}^{p-1}=- P_{ab}(\{p_c,q_{cd*}\})
\end{equation}
where, in matrix form,  $\mathbf P=(\mathbf I + \mathbf U)^{-1}$. All quantities can be expanded in increasing number of free replica sums. 
The saddle-point solution can be expanded as
\begin{equation}
 \label{eq_saddle-point_qab}
\begin{aligned} 
q_{ab*}=q^{[0]}(p_a,p_b)+\sum_c q^{[1]}(p_a,p_b\vert p_c)+\mathcal O(\sum_{cd})
\end{aligned}
\end{equation}
and, similarly, $P_{ab}(\{p_c,q_{cd*}\})=\widehat P_{a*}(\{p_c\})\delta_{ab} + \widetilde P_{ab*}(\{p_c\})$ with
\begin{equation}
\begin{aligned}
 \label{eq_expansion_P}
&\widehat P_{a*}(\{p_c\})=\widehat P^{[0]}(p_a)+\sum_c \widehat P^{[1]}(p_a,p_b\vert p_c)+\cdots
\\& \widetilde P_{ab*}(\{p_c\})=\widetilde P^{[0]}(p_a,p_b)+\sum_c \widetilde P^{[1]}(p_a,p_b\vert p_c)+\cdots
\end{aligned}
\end{equation}
Such expansions in increasing number of free replica sums allow algebraic manipulations that lead to well defined and unique expressions of 
the various orders.\cite{tarjus-tissier,tissier-tarjus, ledoussal-wiese}

The algebra however rapidly becomes cumbersome. An efficient way to proceed, which allows the use of a symbolic software like Mathematica, 
is then as follows. For the calculation of the $k$-th cumulant of the action, one divides the $n$ replicas into $k$ groups of $n_1, \ldots, n_k$ 
replicas having the same overlap $p_1, \ldots, p_k$ with the reference configuration. One then computes the replicated action 
$S_{rep}[p_1, \ldots, p_k]$ corresponding to this ansatz and evaluate the term of order $n_1 \cdots n_k$. In the following we detail how this 
calculation works up to the third cumulant.

Let us start with the first cumulant. We thus consider only one group of replicas, 
namely all $n$ replicas have the same overlap $p_1$ with the reference configuration. As mentioned above, we focus on the local part 
of the cumulant and therefore only consider a spatially uniform overlap field $p_1$. 
Inserting this ansatz into Eq.~(\ref{eq_freereplicasums_S_app}), we obtain
\begin{equation}
\begin{aligned} 
&S_{rep}[p_1]= n S_1[p_1] 
- \frac{n^2}{2} S_2[p_1,p_1] + \frac{n^3}{3!} S_3 [p_1,p_1,p_1] + \ldots 
\end{aligned}
\end{equation}
whereas the saddle-point equation for the overlap $q_1$ between two replicas having overlap $p_1$ with the reference configuration 
[see Eq.~(\ref{eq_saddle-point})] becomes
\begin{equation}
\label{eq_saddle-point}
p\frac{\beta^2}{4} q_{1*}^{p-1}=\frac{q_{1*}-p_1^2}{(1-q_{1*})(1-q_{1*} +n q_{1*}-n p_1^2)}Ê\,.
\end{equation}
As in the conventional replica trick, the first cumulant of the action is simply given by the term of order $n$ of $S_{rep} [p_1]$,
which can be obtained as $\partial_n S_{rep} [p_1] |_{n=0}$.
This object, which only in the mean-field limit coincides with the standard mean-field Franz-Parisi potential,\cite{franzparisi-potential} 
can be computed in a straightforward way,
\begin{equation}
\begin{aligned} 
\label{eq:cumulant1}
L^{-d}S_1(p_1) =- \frac{\beta^2}{4} \left ( 2 p_1^p - q_1^p \right) + \frac{p_1^2 - q_1}{2(1 - q_1)} - \frac{1}{2} \log (1 - q_1),
\end{aligned}
\end{equation}
where $L^d$ is the volume of the system and $q_1$ is given by the saddle point equation, Eq.~(\ref{eq_saddle-point}), when $n\rightarrow 0$ 
(for ease of notation, we drop the star that denotes the saddle-point solution in what follows):
\begin{equation} 
\label{eq:sp1}
\frac{\beta^2 p}{4} [q_1^{(0)}]^{p-1} = \frac{1}{2} \, \frac{q_1^{(0)}-p_1^2}{(1-q_1^{(0)})^2}\, .
\end{equation}

In order to compute the second cumulant of the action, we now consider two groups of replicas, $n_1$ replicas having an overlap $p_1$ 
with the reference configuration, and $n_2$ replicas having overlap $p_2$ with the reference configuration; $q_1$ denotes the overlap 
among replicas of the first group, $q_2$ that among  replicas of the second group, and $q_{12}$ the overlap between a replica of 
the first group and one of the second group.
With this ansatz, Eq.~(\ref{eq_freereplicasums_S_app}) becomes
\begin{equation}
\begin{aligned} 
\label{eq:order2}
&S_{rep}[p_1,p_2]= n_1 S_1[p_1] + n_2 S_1[p_2] \\
& - \frac{1}{2} \left( n_1^2 S_2[p_1,p_1] + n_2^2 S_2[p_2,p_2] + 2 n_1 n_2 S_2[p_1,p_2] \right ) + \ldots
\end{aligned}
\end{equation}
As a result, $S_2[p_1,p_2]$ is simply obtained from the term of order $n_1 n_2$ of $S_{rep}$ and is given by
$-\partial_{n_1,n_2} S_{rep} [p_1,p_2] |_{n_1=n_2=0}$. The saddle-point solutions can also be expanded in powers of 
$n_1$ and $n_2$, \textit{e.g.},
\begin{equation}
\label{eq_expansion_qa}
\begin{aligned}
q_1 & = q_1^{(0)} + n_1 \partial_{n_1} q_1\vert_0 + n_2 \partial_{n_2} q_1\vert_0 + \ldots \\
& = q_1^{(0)} + n_1 q_1^{(10)} + n_2 q_1^{(01)} + \ldots \, ,
\end{aligned}
\end{equation}
where we have introduced a short-hand notation for the derivative of the overlaps with respect to the numbers of replicas (taken 
in the limit $n_1=n_2=0$). From the saddle-point equation for $q_1$, one has for instance
\begin{equation}
\label{eq_first_derivative01}
\begin{aligned}
&\left [p(p-1)\frac{\beta^2}{4} [q_{1}^{(0)}]^{p-2}- \frac{1-2 p_1^2 + q_{1}^{(0)}}{2 (1 - q_{1}^{(0)})^3} \right ]q_1^{(01)}=\\&
\qquad \qquad \qquad \qquad - \frac{(q_{12}^{(0)} - p_1 p_2)^2}{2 (1-q_{1}^{(0)})^2(1-q_{2}^{(0)})}\,.
\end{aligned}
\end{equation}
Contributions to $-\partial_{n_1,n_2} S_{rep} [p_1,p_2] |_{n_1=n_2=0}$ therefore come from the term directly proportional 
to $n_1n_2$ in Eq.~(\ref{eq:order2}) but also from terms that are explicitly linear in $n_1$ or $n_2$ but involve $q_1$ and/or 
$q_2$ that themselves carry a dependence on $n_1$ and $n_2$ at the saddle-point [see Eq.~(\ref{eq_expansion_qa})]. 

After some manipulations, one then finds
\begin{equation} \label{eq:cumulant2}
\begin{aligned} 
L^{-d}S_2[p_1,p_2] =  \frac{\beta^2}{2} [q_{12}^{(0)}]^p - \frac{(q_{12}^{(0)} - p_1 p_2)^2}{2 (1 - q_1^{(0)}) (1 - q_2^{(0)})},
\end{aligned}
\end{equation}
where $q_{12}^{(0)}$ satisfies the saddle-point equation at this order,
\begin{equation}
\label{eq:sp12}
\frac{\beta^2 p}{2} [q_{12}^{(0)}]^{p-1} = \frac{q_{12}^{(0)}- p_1 p_2}{(1 - q_1^{(0)}) (1 - q_2^{(0)})} \,,
\end{equation}
and the contributions involving the first-order derivatives of $q_1$ and $q_2$ (see above) exactly cancel due to the saddle-point condition.

For computing the third cumulant, we divide the $n$ replicas in 
three groups of $n_1$, $n_2$, and $n_3$ replicas, having an overlap  with the reference configuration 
equal to $p_1$, $p_2$, and $p_3$,  respectively; $q_1,q_2,q_3$ are the overlaps inside a given group and $q_{12}, q_{13}, q_{23}$ 
the overlaps among replicas belonging to distinct groups. As for the first and second cumulants,  
$S_3[p_1,p_2,p_3]$ is simply given by the term of order $n_1 n_2 n_3$ of $S_{rep}$, and can be obtained as
$\partial_{n_1,n_2,n_3} S_{rep} [p_1,p_2,p_3] |_{n_1=n_2=n_3=0}$.
There are several contributions to this term, coming from the formal 
expression of $\mathcal S[\{p_a,q_{ab}\}]$ when three groups of replicas are introduced (considering again uniform overlaps): 
\begin{equation}
\begin{aligned} 
\label{eq:order3}
&L^{-d}\mathcal S[\{p_1,p_2,p_3,q_1,q_2,q_3,q_{12},q_{13},q_{23}\}]=\\& \sum_{a=1}^3 n_a A_a - \frac{1}{2} \sum_{a,b=1}^3 n_a n_b B_{ab}+ 
\frac{1}{3!} \sum_{a,b,c=1}^3 n_a n_b n_c C_{abc}+ \ldots
\end{aligned}
\end{equation}
where $A_a \equiv A(p_a, q_a)$,  $B_{ab}\equiv B(p_a, p_b,q_a,q_b,q_{ab})$, and 
$C_{abc}\equiv C(p_a, p_b,p_c, q_a,q_b,q_c,q_{ab},q_{bc},q_{ca})$ (with the convention $q_{aa}=q_a$ when $b=a$).

The simplest term is that having an explicit dependence on $n_1n_2n_3$. It is given by
\begin{equation}
\begin{aligned}
\label{eq_C}
& C(p_1,p_2,p_3,q_1,q_2,q_3,q_{12},q_{13},q_{23}) \\ 
& \qquad = -\frac{(q_{12} - p_1 p_2)(q_{13} - p_1 p_3)(q_{23} - p_3 p_3)}
{(1 - q_1) (1 - q_2) (1 - q_3)} \,,
\end{aligned}
\end{equation}
where all overlaps can be taken at the lowest order, \textit{i.e.}, $q_1=q_1^{(0)}$, $q_{12}=q_{12}^{(0)}$, etc. 

The second contribution is given by the expansion of the terms of order $n_1 n_2$, $n_1 n_3$, and $n_2 n_3$ of $S_{rep}$ to first order
in, respectively, $n_3$, $n_2$, and $n_1$ around the saddle-point values of the overlaps:
\begin{equation}
\begin{aligned}
& - \left( \partial_{q_1} B_{12} \right) q_1^{(001)} - \left( \partial_{q_2} B_{12} \right) q_2^{(001)} 
- \left( \partial_{q_{12}} B_{12} \right) q_{12}^{(001)} \\
& - \left( \partial_{q_1} B_{13} \right) q_1^{(010)} - \left( \partial_{q_3} B_{13} \right) q_2^{(010)} 
- \left( \partial_{q_{13}} B_{13} \right) q_{13}^{(010)} \\
& - \left( \partial_{q_2} B_{23} \right) q_2^{(100)} - \left( \partial_{q_3} B_{23} \right) q_3^{(100)} 
- \left( \partial_{q_{23}} B_{23} \right) q_{23}^{(100)},
\end{aligned}
\end{equation}  
where the derivatives are taken for $n_1=n_2=n_3=0$ and the notation $q_1^{(001)}$, $q_{12}^{(001)}$, etc, is an obvious generalization 
to three groups of replicas of that introduced above for two groups (and are therefore given by expressions similar to 
Eq.~(\ref{eq_first_derivative01}). $B_{ab}$ is expressed as 
\begin{equation}
\begin{aligned}
\label{eq_B}
B(p_a,p_b,q_a,q_b,q_{ab}) =  \frac{\beta^2}{2} q_{ab}^p - \frac{(q_{ab} - p_a p_a)^2}{2 (1 - q_a) (1 - q_b)} \,.
\end{aligned}
\end{equation}

The third and last contribution comes from the expansion of the terms of order $n_1$, $n_2$, and $n_3$ of $S_{rep}$ up to
second order in, respectively, $n_2 n_3$, $n_1 n_3$, and $n_1 n_2$ around the saddle-point values of the overlaps:
\begin{equation}
\begin{aligned}
& \left( \partial_{q_1} A_1 \right) q_1^{(011)} 
+ \left( \partial_{q_2} A_2 \right) q_2^{(101)}  
+ \left( \partial_{q_3} A_3 \right) q_3^{(110)} \\
& + \left( \partial^2_{q_1} A_1 \right) q_1^{(010)} q_1^{(001)}
+ \left( \partial^2_{q_2} A_2 \right) q_2^{(100)} q_2^{(001)} \\
& + \left( \partial^2_{q_3} A_3 \right) q_3^{(100)} q_3^{(010)}. 
\end{aligned}
\end{equation} 

After combining the three contributions and using the the saddle-point equations for the overlaps, Eq.~(\ref{eq:sp1}) and Eq.~(\ref{eq:sp12}), 
several terms cancel out and the expression of the third cumulants simplifies to
\begin{equation} \label{eq:cumulant3}
\begin{aligned}
S_3 [p_1,p_2,p_3] & = C_{123}^{(0)} - \left( \partial_{q_2} B_{12} \right) q_2^{(001)} \\
& - \left( \partial_{q_1} B_{13} \right) q_1^{(010)}
- \left( \partial_{q_3} B_{23} \right) q_3^{(100)}\,,
\end{aligned}
\end{equation}
where $C_{123}^{(0)}$ is given by Eq.~(\ref{eq_C}) with all overlaps given by their zeroth-order expression, $B_{ab}$ is given by Eq.~(\ref{eq_B}), 
and $q_1^{(010)}$ and related terms are given by expressions similar to Eq.~(\ref{eq_first_derivative01}).

The calculation of the fourth cumulant can be carried out along similar lines by introducing 4 groups of replicas. For simplicity, 
we do not write down its explicit expression, which can be easily obtained with Mathematica.

With the explicit expressions of the cumulants of the action in hand, one can easily compute the values of the coefficients of
the different terms at the terminal critical point of the transition line in the $T$-$\epsilon$ diagram. We illustrate this for the case 
$p=3$.
The first step is to determine the location of the terminal point at the saddle-point level. 
By imposing that $\partial^2_{p_1} S_1[p_1] = \partial^3_{p_1} S_1[p_1] = 0$ and using Eq~(\ref{eq:sp1}), we find that $p_{1,c}= 0.285$, 
$\beta_c = 1.295$, and $q_{1,c} = q_{12,c} = 0.103$.
We next expand the overlaps with the reference configuration around the terminal critical point, $p_a = p_{1,c} + \phi_a$, and compute
all the coefficients of the expansion of the replicated action $S_{rep}[\{p_a\}]$ up to fourth order in the $\phi_a$'s. 
The notations are the same as in the main text and we obtain: $r_4= 55.05,\Delta_{20}=0.18,\Delta_{21}=1.55,\Delta_{22}=15.45,
\Delta_{23}=2.19,\Delta_{30}=0.36,\Delta_{31}=7.18,\Delta_{40}=5.27$. 
As for the Landau-like functional considered in the main text, the important point is that $r_4$, 
$\Delta_{20}$, $\Delta_{22}$, and $\Delta_{40}$ are greater than zero. 
Finally, we have checked that the eigenvalues of the Hessian at the saddle point are strictly positive:  $\lambda_R=0.052$ and  
$\lambda_{L,A}=0.067$. 
This guarantees the consistency of the whole procedure. \\The procedure for the Landau-like functional considered in the main text
is analogous (the computation is slightly simpler). All the results are quoted in the main text.

\section{Surface tension and bare disorder strength in the effective RFIM}

The conclusion of the analysis presented in the main text is that the terminal critical point found in the presence of a conjugate source within mean-field theory belongs 
in finite dimension to the universality class of the RFIM. This is valid {\it if the transition is not destroyed by the disorder}. 
In order to assess whether the transition exists when all fluctuations beyond mean-field theory are taken into account we 
have compared the ``bare" strength of the random field, $\sqrt \Delta$, obtained from $S_2$ to that of the surface tension $Y$ obtained from $S_1$. 
Clearly, the location of terminal critical point found within mean-field theory is changed by fluctuations. The question is whether 
it is just shifted down to lower temperature 
or if the transition is wiped out altogether. In the following we present the computation for the Landau-like functional.\cite{Dzero-Schmalian-Wolynes} \\
We have computed $Y$ as the free energy cost per unit surface between two regions with high and low overlap\cite{langer,franz} 
at coexistence far below the (mean-field) 
critical point. We have chosen to evaluate $Y$ at the (mean-field) Kauzmann temperature $T_K$; the idea is that if the disorder is not able to renormalize $Y$ 
to zero, and hence to destroy the first-order transition close to $T_K$, then a terminal critical point exists for sure.\cite{footnote1}\\
The computation of $Y$ is based on a standard instanton computation \cite{langer,franz}. Let us rewrite $S_1$ as:
\[
 S_1[p_a]=\frac{E_0}{k_BT}\int_x \big [c (\partial p_a(x))^2+\tilde{V}(p_a(x))\big]
\]
where 
\begin{eqnarray}
\tilde{V}(p_a)&&=2V(p_a)-V(q^{[0]}(p_a,p_a))\\
&&-\frac{2u}{3}q^{[0]}(p_a,p_a)^3+u p_a^2q^{[0]}(p_a,p_a)\nonumber
\end{eqnarray} 
and, for simplicity, we have neglected
the contribution $-\frac 1 2 c (\partial q^{[0]}(p_a,p_a))^2$ to the kinetic term (at least close to the terminal critical point it leads to a very small correction). 
The expression for the bare surface tension obtained by an instanton computation at $T_K$ is:
\[
Y=\frac{2E_0}{k_BT_K}\sqrt{c}\int_0^{q_{ea}}\sqrt{\tilde{V}(p)}dp
\]
where $0$ and $q_{ea}$ are the values of $p$ corresponding to the two minima of $\tilde{V}(p)$ and $E_0/(k_BT_K)=3.57$ for the values of the parameters 
chosen to mimic liquid otho-terphenyl\cite{Dzero-Schmalian-Wolynes} ($q_{ea}$ is equal to 1 in this case). 
By computing $\tilde{V}(p)$ at $T_K$ (this requires breaking replica symmetry for $q_{ab}$, see [\onlinecite{franzparisi-potential}]) and taking $c=1$ 
we have obtained $Y\simeq 1.02$. 
When repeating the analysis for the $p$-spin  model with $p=3$, we have found $Y\simeq 0.095$.

These values must be compared to the strength of the disorder, which has of course to be computed at the same temperature $T_K$. 
Because of the quenched disorder the two minima at $p=0$ and $p=q_{ea}$ have the same height only in average. The height fluctuations can 
destroy the long-range order found within mean field theory even in the absence of thermal noise, as it is well known from the theory of the random field 
Ising model (RFIM).\cite{nattermann} In order to evaluate the strength of these fluctuations we compute for uniform values of the overlap field 
\[
\Delta=\frac 1 4 L^{-d} \, \overline{(S[0\vert {\bf{r}_0^N}]- S[q_{ea}\vert {\bf{r}_0^N}])^2} \,.
\]
When the overlap field reduces to an Ising variable, {\it i.e.} it can just take the two values $0$ and $q_{ea}$, which seems a reasonable approximation at 
$T_K$, the above expression indeed coincides with the 
variance of the random field. By using the formalism discussed in the main text we find
\[
\Delta=\frac 1 4 L^{-d} (S_2[0,0]+S_2[q_{ea},q_{ea}]-2S_2[0,q_{ea}])\,.
\]
This expression further simplifies further since $S_2[0,0]=S[0,q_{ea}]=0$. A simple computation then leads to 
$\Delta=\frac{E_0}{k_BT_K}u q_{ea}^3/6 \simeq 0.23$. 
(recall that $q_{ea}=1$ in this case).  Repeating the analysis for the 3-spin model then gives $\Delta=q_{ea}^2 (2/(3-3q_{ea})-1)/8\simeq 0.045$ 
($q_{ea}=0.644$ in this case).\\
From the above values we derive the ratios $\sqrt{\Delta}/Y\simeq 0.47$ for the Landau functional and $\sqrt{\Delta}/Y\simeq 2.24$ for the 
3-spin model quoted in the main text. The critical value of the disorder for the RFIM on a cubic lattice is $\sqrt{\Delta}/J\vert_c\simeq 2.3$, where $J$ 
is the coupling interaction between spins.\cite{nattermann,middleton-fisher} 
If the glass-forming system at $T_K$ can be considered as a hard Ising spin model, then $Y$ is twice the coupling $J$. A transition is therefore expected for 
the Landau-like functional for ortho-terphenyl but not for the 3-spin model.


\begin{thebibliography}{99}
\bibitem{footnote}
We have explicitly verified this point by computing the eigenvalues of the Hessian operator: when $n\rightarrow 0$ there are two 
strictly positive eigenvalues: $\lambda_R=1.315$ and  $\lambda_{L,A}=1.370$.
\end{thebibliography}
\begin{thebibliography}{99}
\bibitem{footnote0}
Compared to the case studied in \cite{franz,franz05} we have chosen a simpler kinetic term which is quadratic in the replica field. 
We do not expect this modification to alter our results.
\bibitem{footnote1}
Note that except at $T=T_K$ the transition is a conventional first-order one and our whole analysis applies.
\end{thebibliography}
\end{document}